\documentclass[aps,prd,preprint,preprintnumbers,nofootinbib,superscriptaddress]{revtex4}
\usepackage{ae}
\usepackage{bm} 
\usepackage{color}
\usepackage{amsmath}
\usepackage{amssymb}
\usepackage{amsfonts}
\usepackage{graphicx}
\usepackage{slashed}
\usepackage{setspace}
\usepackage{stackrel}
\usepackage{ulem}

\newcommand{\Eqref}[1]{Eq.~\eqref{#1}}

\allowdisplaybreaks

\begin{document}

\setlength{\unitlength}{1mm}
\title{The Heisenberg-Euler effective action in slowly varying electric field inhomogeneities of Lorentzian shape}
\author{Felix Karbstein}\email{felix.karbstein@uni-jena.de}
\affiliation{Helmholtz-Institut Jena, Fr\"obelstieg 3, 07743 Jena, Germany}
\affiliation{Theoretisch-Physikalisches Institut, Abbe Center of Photonics, \\ Friedrich-Schiller-Universit\"at Jena, Max-Wien-Platz 1, 07743 Jena, Germany}

\date{\today}

\begin{abstract}
 We use a locally constant field approximation (LCFA) to study the one-loop Heisenberg-Euler effective action in a particular class of slowly varying inhomogeneous electric fields of Lorentzian shape with $0\leq d\leq 4$ inhomogeneous directions.
 We show that for these fields, the LCFA of the Heisenberg-Euler effective action can be represented in terms of a single parameter integral, with the constant field effective Lagrangian with rescaled argument as integration kernel.
 The imaginary part of the Heisenberg-Euler effective action contains information about the instability of the quantum vacuum towards the formation of a state with real electrons and positrons.
 Here, we in particular focus on the dependence of the instantaneous vacuum decay rate on the dimension $d$ of the field inhomogeneity.
 Specifically for weak fields, we find an overall parametric suppression of the effect with $(E_0/E_{\rm cr})^{d/2}$, where $E_0$ is the peak field strength of the inhomogeneity and $E_{\rm cr}$ the critical electric field strength. 
\end{abstract}

\maketitle

\section{Introduction}

The quantum vacuum in external electromagnetic fields has peculiar properties, cf. the pertinent reviews \cite{Fradkin:1991zq,Dittrich:1985yb,Dittrich:2000zu,Dunne:2004nc,Dunne:2008kc} and references therein.
The response of the quantum vacuum to an external electromagnetic field has been formalized in the Heisenberg-Euler effective action \cite{Heisenberg:1935qt,Schwinger:1951nm}, which was first derived by W.~Heisenberg and H.~Euler more than eighty years ago for constant electromagnetic fields, and at one-loop order in the fluctuating quantum fields. More specifically, in the language of Feynman diagrams, the one-loop Heisenberg-Euler action amounts to a virtual electron positron loop, taking into account the couplings of the virtual electrons/positrons to the external field to all orders.
It encodes quantum corrections to Maxwell's classical theory of electromagnetic fields in terms of nonlinear interactions between external electromagnetic fields.

Apart from the constant-field results, only a few exact results for the one-loop Heisenberg-Euler action in specific one-dimensional field inhomogeneities are known explicitly; cf., e.g., \cite{Cangemi:1994by,Cangemi:1995ee,Dunne:1998ni,Dunne:1997kw,Kim:2009pg}, and \cite{Dunne:2004nc,Dunne:2008kc} for reviews.
In particular, no analytical results are known for field inhomogeneities with $d>1$ inhomogeneous directions.

In this article, we adopt a locally constant field approximation (LCFA) to the one-loop Heisenberg-Euler action. By limiting ourselves to a specific class of electric field profiles, namely electric fields of Lorentzian shape with $0\leq d\leq4$ inhomogeneous directions, we obtain explicit analytical insights into the corresponding effective action for slowly varying fields. Our results are also applicable for purely magnetic fields of the same field profile as the effective actions in purely electric and magnetic fields are related by an electric-magnetic duality, cf., e.g., \cite{Jentschura:2001qr}. 

One of the most striking consequences of quantum vacuum nonlinearities in external electric fields is the instability of the quantum vacuum towards a state with real electrons and positrons.
The corresponding phenomenological signature would be electron-positron pair-creation from vacuum.
On the level of the Heisenberg-Euler effective action this effect manifests itself in an non-zero imaginary part.
In constant electric fields of strength $E$, this imaginary part, and thus the instantaneous vacuum decay rate can be determined exactly, resulting in the renowned Schwinger formula \cite{Heisenberg:1935qt,Schwinger:1951nm}.
The vacuum decay rate becomes sizable only for electric field strengths of the order of the critical electric field strength $E_{\rm cr}\equiv\frac{m^2c^3}{e\hbar}\simeq1.3\cdot10^{18}{\rm V}/{\rm m}$ \cite{Sauter:1931zz,Heisenberg:1935qt,Schwinger:1951nm}, due to a nonperturbative exponential suppression $\sim\exp\{-\pi E_{\rm cr}/E\}$.
The critical electric field strength amounts to the electric field strength acquired by an electron (charge $e$, mass $m$) over the distance of its Compton wavelength $\lambdabar_{\rm C}=\frac{\hbar}{mc}=3.86 \cdot 10^{-13}{\rm m}$.
Semi-classically, it can be interpreted as the field strength needed to promote a virtual electron-positron pair, probing distances of the order of the Compton wavelength of the electron, to real on-shell particles.

The imaginary part of our LCFA result for the Heisenberg-Euler action in electric field profiles of Lorentzian shape with $0\leq d\leq4$ inhomogeneous directions allows us to study the generalization of Schwinger's formula in these slowly varying fields.

\section{Results}

The on-shell renormalized, one-loop Heisenberg-Euler effective Lagrangian in a uniform constant electric field $\vec{E}$ has the propertime representation \cite{Heisenberg:1935qt,Schwinger:1951nm},
\begin{equation}
 {\cal L}^{1\text{-loop}}_{\rm HE}(E)=\frac{m^4}{8\pi^2}\int_0^\infty\frac{{\rm d}s}{s}\,\Bigl(\frac{eE}{m^2}\frac{1}{s}\Bigr)^2{\rm e}^{-{\rm i}\frac{m^2}{eE}s}\,
 \biggl\{s\coth(s)-\frac{1}{3}s^2-1\biggr\}\,, \label{eq:L1}
\end{equation}
where $E=|\vec{E}|$, and the prescription $m^2\equiv m^2-{\rm i}0^+$ is implicitly understood throughout this article.
For an detailed discussion of the emergence of the Heisenberg-Euler effective action from the microscopic theory of quantum electrodynamics (QED), we refer the reader to \cite{Gies:2016yaa}, and references therein.
We use the Heaviside-Lorentz System with units where $c=\hbar=1$.
The interesting feature of \Eqref{eq:L1} is that the entire dependence on the electric field can be encoded in the function
\begin{equation}
 f\bigl(\tfrac{eE}{m^2}\tfrac{1}{s}\bigr):=\Bigl(\frac{eE}{m^2}\frac{1}{s}\Bigr)^2{\rm e}^{-{\rm i}\frac{m^2}{eE}s} \label{eq:f}
\end{equation}
in the integrand of the propertime integral.
As the effective Lagrangians in purely electric and magnetic fields are related by an electric-magnetic duality, cf., e.g., \cite{Jentschura:2001qr,Karbstein:2013ufa},
the analogous expression in a purely magnetic field follows by the substitution $E\to{\rm i}B$.

Sticking to a locally constant field approximation (LCFA) of the effective Lagrangian \cite{Galtsov:1982,Karbstein:2015cpa,Gies:2016yaa,Gavrilov:2016tuq},
\begin{equation}
 {\cal L}_\text{HE}(E) \to {\cal L}_\text{HE}\bigl(E(x)\bigr)\,, \label{eq:LCFA}
\end{equation}
we can use the result~\eqref{eq:L1} for constant electric fields to obtain insights into slowly varying, spatially inhomogeneous electric fields.
In turn, the corresponding effective action is a functional of $E(x)$ and reads $\Gamma_\text{HE}\bigl[E(x)\bigr]=\int{\rm d}^4x\,{\cal L}_\text{HE}\bigl(E(x)\bigr)$.
The deviations of this LCFA result from the corresponding -- typically unknown -- exact result for $\Gamma_\text{HE}$ in the particular inhomogeneous background field profile under consideration are
of order ${\cal O}\bigl((\tfrac{\upsilon}{m})^2\bigr)$, where $\upsilon$ delimits the moduli of the frequency and momentum components of the considered inhomogeneous field from above \cite{Galtsov:1982,Karbstein:2015cpa}. Hence, the LCFA amounts to keeping the leading term in an expansion in $\tfrac{\upsilon}{m}\to0$.
Generically, the results obtained from a LCFA should be reliable for inhomogeneous fields which vary on spatial scales much larger than the Compton wavelength of the electron $\lambdabar_{\rm C}$. However, the precise regime of its applicability is not a priori clear and very much depends on the other (potentially competing) dimensionless parameters governing the problem under consideration, cf. also \cite{Dunne:1999uy}.
Note that the LCFA is equivalent to the zeroth-order term of a derivative expansion of the effective action, cf., e.g., \cite{Dunne:1998ni}.
For dedicated studies of higher-order contributions in the derivative expansion of the Heisenberg-Euler effective action, cf. \cite{Cangemi:1994by,Gusynin:1995bc,Gusynin:1998bt,Dunne:1999uy,Dunne:2000up}.

In this article, we employ the LCFA~\eqref{eq:LCFA} for a specific class of spatially inhomogeneous fields of Lorentzian profile,
\begin{equation}
 \vec{E}(x)=E(x)\,\hat{\vec{e}}_E\,,\quad\text{with}\quad E(x):=\frac{E_0}{1+\sum_{i=1}^d(\frac{2x_i}{w_i})^2}\,, \label{eq:E(x)}
\end{equation}
where $0\leq d\leq 4$ counts the inhomogeneous directions of the field, and $w_i$ is the full width at half maximum in direction $\hat{\vec{e}}_i$.
The electric field vector points in a fixed direction $\hat{\vec{e}}_E$, which does not depend on the space-time coordinates.
In four space-time dimensions, a field profile with $d$ inhomogeneous directions is of course homogeneous in the remaining $4-d$ directions.
Let us also remark, that for the sake of a transparent notation, the inhomogeneous field profile~\eqref{eq:E(x)} is represented in such a way, that all the potentially inhomogeneous directions labeled by the index $i$ amount to spatial directions.  
However, the calculations performed here remain valid if one of the $d$ directions of the inhomogeneity is identified with time, such that $\frac{2x_j}{w_i}\to\frac{2t}{T}$ for one $j\in\{1,\ldots d\}$, where $T$ is the pulse duration.
Other possibilities include, e.g., a light-like dependence of the form $\frac{2x_j}{T}\to\frac{2(x_j-t)}{T}$.

Let us emphasize once again that as a consequence of the electric-magnetic duality (cf. above), by means of the substitution $E_0\to{\rm i}B_0$ all the results derived below for inhomogeneous electric fields of peak amplitude $E_0$ can be straightforwardly adopted for the corresponding purely magnetic fields of peak amplitude $B_0$.
Recall, however, that the Heisenberg-Euler effective action does not have an imaginary part in purely magnetic fields.

The field profile~\eqref{eq:E(x)} is special in the sense that for finite widths $w_i$ in the inhomogeneous directions and for $d>0$, it fulfills the following identity (cf. appendix~\ref{app:id} for its derivation),
\begin{equation}
 \int{\rm d}^4x\,f\bigl(\tfrac{eE(x)}{m^2}\tfrac{1}{s}\bigr)=V_{(4-d)}\biggl(\prod_{i=1}^d w_i\biggr)
 \Bigl(\frac{\pi}{4}\Bigr)^{\frac{d}{2}}\frac{1}{\Gamma(\frac{d}{2})}
 \int_0^\infty\frac{{\rm d}u}{u}\,u^{\frac{d}{2}} f\bigl(\tfrac{eE_0}{m^2}\tfrac{1}{s(1+u)}\bigr)\,, \label{eq:Erep}
\end{equation}
with $f(.)$ as defined in \Eqref{eq:f}.
The integration over the $4-d$ homogeneous directions is trivial and results in an overall volume factor $V_{(4-d)}:=\int{\rm d}^{4-d}x$.

Equation~\eqref{eq:Erep} implies that within the LCFA, the effective action in the field~\eqref{eq:E(x)} with $d>0$ can be represented as
\begin{equation}
 \Gamma_\text{HE}^{1\text{-loop}}\bigl[E(x)\bigr]=V_{(4-d)}\biggl(\prod_{i=1}^d w_i\biggr)
 \Bigl(\frac{\pi}{4}\Bigr)^{\frac{d}{2}}\frac{1}{\Gamma(\frac{d}{2})}
 \int_0^\infty\frac{{\rm d}u}{u}\,u^{\frac{d}{2}}\,{\cal L}_\text{HE}^{1\text{-loop}}\bigl(\tfrac{E_0}{1+u}\bigr)\,, \label{eq:GammaHE(x)}
\end{equation}
while, of course, $\Gamma_\text{HE}^{1\text{-loop}}\bigl[E(x)\bigr]=V_{(4)}{\cal L}_\text{HE}^{1\text{-loop}}(E_0)$ for $d=0$. 
From the structure of \Eqref{eq:GammaHE(x)}, it is obvious that insights into inhomogeneous electric fields~\eqref{eq:E(x)} with peak field strength $E_0$ require knowledge about ${\cal L}_\text{HE}^{1\text{-loop}}(E)$ in the entire field strength regime $0\leq E\leq E_0$.
Hence, for weak peak field strengths $E_0\ll\frac{m^2}{e}$, the evaluation of the integrand in \Eqref{eq:GammaHE(x)} only requires knowledge about the weak field regime of ${\cal L}_\text{HE}^{1\text{-loop}}(E)$.
With the help of \eqref{eq:IntE2n(x)} it is straightforward to generalize the perturbative expansion of the Heisenberg-Euler Lagrangian in constant electric fields \cite{Dunne:2004nc}
to the inhomogeneous field~\eqref{eq:E(x)}, resulting in
\begin{multline}
 \Gamma_\text{HE}^{1\text{-loop}}\bigl[E(x)\bigr]\sim V_{(4-d)}\biggl(\prod_{i=1}^d w_i\biggr)\Bigl(\frac{\pi}{4}\Bigr)^{\frac{d}{2}}\frac{m^4}{8\pi^2} \\
 \times\sum_{n=0}^\infty\frac{\Gamma(2n+4-\frac{d}{2})}{\Gamma(2n+4)}\,\frac{(-1)^{n+1}\,{\cal B}_{2n+4}}{(2n+4)(2n+3)(2n+2)}\Bigl(\frac{2eE_0}{m^2}\Bigr)^{2n+4}\,, \label{eq:perts}
\end{multline}
valid for $d\geq0$, with Bernoulli numbers ${\cal B}_{2n}$.

This is different for strong $E_0\gtrsim\frac{m^2}{e}$, where ${\cal L}_\text{HE}^{1\text{-loop}}\bigl(\tfrac{E_0}{1+u}\bigr)$ generically interpolates between strong and weak field limits of ${\cal L}_\text{HE}^{1\text{-loop}}(E)$ as a function of $u$.
Thus, to obtain the strong field limit of \Eqref{eq:GammaHE(x)} it does not suffice to simply adopt the strong field expansion of ${\cal L}_\text{HE}^{1\text{-loop}}(E)$ in the integrand of \Eqref{eq:GammaHE(x)}.
Note that the strong field expansion of ${\cal L}_\text{HE}^{1\text{-loop}}(E)$ is straightforwardly obtained from Eq.~(1.62) of \cite{Dunne:2004nc} by the substitution $B\to-{\rm i}E$ (cf. above).
A naive integration of this expression in the integrand of \Eqref{eq:GammaHE(x)} would even result in divergent contributions arising from large $u$, for which the assumptions invoked to arrive at this expansion are manifestly violated.

Remarkably, for constant electric fields of arbitrary strength the propertime integral in \Eqref{eq:L1} can even be performed explicitly \cite{Dittrich:1975au,Dittrich:1985yb,Dunne:2004nc}, resulting in the closed-form representation
\begin{multline}
 {\cal L}_\text{HE}^\text{1-loop}(E)=\frac{m^4}{8\pi^2}\biggl\{\frac{1}{3}\Bigl(\frac{eE}{m^2}\Bigr)^2
 \biggl[1+\ln\Bigl(\frac{\rm i}{2}\frac{m^2}{eE}\Bigr)-12\zeta'\Bigl(-1,\frac{\rm i}{2}\frac{m^2}{eE}\Bigr)\biggr] \\
 -{\rm i}\frac{eE}{m^2}\ln\Bigl(\frac{\rm i}{2}\frac{m^2}{eE}\Bigr)+\frac{1}{4}\biggl[1-2\ln\Bigl(\frac{\rm i}{2}\frac{m^2}{eE}\Bigr)\biggr]\biggr\}\,, \label{eq:L2}
\end{multline}
where $\zeta'(t,\chi)=\partial_t\zeta(t,\chi)$ denotes the first derivative of the Hurwitz zeta function.
In turn, we have the following parameter integral representation of the LCFA effective action in the background field~\eqref{eq:E(x)} with $d>0$, 
\begin{multline}
 \Gamma_\text{HE}^{1\text{-loop}}\bigl[E(x)\bigr]=V_{(4-d)}\biggl(\prod_{i=1}^d w_i\biggr)
 \Bigl(\frac{\pi}{4}\Bigr)^{\frac{d}{2}}\frac{m^4}{8\pi^2}\frac{1}{\Gamma(\frac{d}{2})}
 \int_0^\infty\frac{{\rm d}u}{u}\,u^{\frac{d}{2}} \\
 \times\biggl\{\frac{1}{3}\Bigl(\frac{eE_0}{m^2(1+u)}\Bigr)^2
 \biggl[1+\ln\Bigl(\frac{\rm i}{2}\frac{m^2(1+u)}{eE_0}\Bigr)-12\zeta'\Bigl(-1,\frac{\rm i}{2}\frac{m^2(1+u)}{eE_0}\Bigr)\biggr] \\
 -{\rm i}\frac{eE_0}{m^2(1+u)}\ln\Bigl(\frac{\rm i}{2}\frac{m^2(1+u)}{eE_0}\Bigr)+\frac{1}{4}\biggl[1-2\ln\Bigl(\frac{\rm i}{2}\frac{m^2(1+u)}{eE_0}\Bigr)\biggr]\biggr\}\,,
\end{multline}
which can be adopted to study the behavior of $\Gamma_\text{HE}^{1\text{-loop}}\bigl[E(x)\bigr]$ for arbitrary values of $E_0$.

In the remainder of this article we focus on the imaginary part of the Heisenberg-Euler effective action in the electric field inhomogeneity~\eqref{eq:E(x)},
which encodes information about the process of electron-positron pair creation from the quantum vacuum in the external electric field \cite{Sauter:1931zz,Heisenberg:1935qt,Schwinger:1951nm,Nikishov:1969tt}.
The well-known result in constant electric fields is given by the {\it Schwinger formula} \cite{Schwinger:1951nm},
\begin{equation}
 {\rm Im}\bigl\{{\cal L}^{1\text{-loop}}_{\rm HE}(E)\bigr\}
 =\frac{(eE)^2}{8\pi^3}\,\sum_{n=1}^\infty \frac{1}{n^2}\, {\rm e}^{-\frac{m^2}{eE}\pi n}\,, \label{eq:ImL1}
\end{equation}
from which we infer the following LCFA result in the inhomogeneous field~\eqref{eq:E(x)},
\begin{align}
 {\rm Im}\bigl\{\Gamma_\text{HE}^{1\text{-loop}}\bigl[E(x)\bigr]\bigr\}&=V_{(4-d)}\biggl(\prod_{i=1}^d w_i\biggr)
 \Bigl(\frac{\pi}{4}\Bigr)^{\frac{d}{2}}\frac{1}{\Gamma(\frac{d}{2})}
 \int_0^\infty\frac{{\rm d}u}{u}\,u^{\frac{d}{2}}\,{\rm Im}\bigl\{{\cal L}_\text{HE}^{1\text{-loop}}\bigl(\tfrac{E_0}{1+u}\bigr)\bigr\} \nonumber\\
 &=V_{(4-d)}\biggl(\prod_{i=1}^d w_i\biggr)
 \Bigl(\frac{\pi}{4}\Bigr)^{\frac{d}{2}}\frac{(eE_0)^2}{8\pi^3} \nonumber\\
 &\quad\quad\quad\times\sum_{n=1}^\infty \frac{1}{n^2}\biggl[\Gamma\Bigl(2-\frac{d}{2},\frac{m^2}{eE_0}n\pi\Bigr)-\frac{m^2}{eE_0}n\pi\,\Gamma\Bigl(1-\frac{d}{2},\frac{m^2}{eE_0}n\pi\Bigr)\biggr]\,, \label{eq:ImGammaHE(x)}
\end{align}
where $\Gamma(a,\chi)$ is the incomplete gamma function. The final expression in \Eqref{eq:ImGammaHE(x)} holds for $d\geq0$.
To perform the integration over the parameter $u$ in the above expression, we employed \Eqref{eq:s2} reversely and made use of formula 3.381.3 of \cite{Gradshteyn}.

A semiclassical viewpoint suggests that the creation of real electron-positron pairs requires the electric field to be sufficiently strong or extended to provide an electrostatic energy greater than the rest energy of the pair \cite{Gies:2015hia}.
This results in the additional constraint,
\begin{equation}
 e\int_{-\infty}^\infty{\rm d}\ell\,E(\ell e_E)\stackrel{!}{>}2m\,, \label{eq:constraint}
\end{equation}
to be necessarily fulfilled for vacuum decay and, thus, pair creation to become possible. 
In \Eqref{eq:constraint}, the electric field profile $E(x)$ in \Eqref{eq:E(x)} is evaluated at $x^\mu=\ell e_E^\mu=\ell(0,\hat{\vec{e}}_E)$, such that $l$ parameterizes the spatial coordinate along the direction of the electric field vector $\hat{\vec{e}}_E$.
Introducing the effective length of the inhomogeneity along the electric field as $l_E:=\frac{1}{E_0}\int_{-\infty}^\infty{\rm d}\ell\,E(\ell e_E)$, cf. also \Eqref{eq:IntE(x)},
the criterion~\eqref{eq:constraint} can be written as
\begin{equation}
 \gamma:=\frac{1}{ml_E}\Bigl(\frac{eE_0}{m^2}\Bigr)^{-1} \stackrel{!}{<}\frac{1}{2}\,. \label{eq:constraint1}
\end{equation}
From \Eqref{eq:constraint1} it is particularly obvious why it is not possible to arrive at the above criterion within the LCFA~\eqref{eq:LCFA}, which manifestly neglects terms suppressed by inverse powers of the dimensionless ratio $ml_E=l_E/\lambdabar_{\rm C}$, cf. above.
The importance of the Keldysh-type parameter $\gamma$ for the vacuum decay process in inhomogeneous fields suggests that the LCFA of the manifestly nonperturbative imaginary part of the Heisenberg-Euler effective action allows for trustworthy insights only as long as $\gamma\ll1$. This reasoning is substantiated by explicit results for $d=1$, cf., e.g., \cite{Dunne:2006st}.
For a detailed discussion of the dimensionless expansion parameters in the case of $d=1$, see \cite{Dunne:1999uy}.

Using the asymptotic expansion of $\Gamma(a,\chi)$ for $a$ fixed and $\chi\to\infty$ \cite{dlmf:1} in \Eqref{eq:ImGammaHE(x)}, we obtain the following weak field expansion:
\begin{multline}
 {\rm Im}\bigl\{\Gamma_\text{HE}^{1\text{-loop}}\bigl[E(x)\bigr]\bigr\}=\biggl(\prod_{i=1}^d w_i\biggr)
 \int{\rm d}^{4-d}x\,\frac{(eE_0)^2}{8\pi^3}\,\Bigl(\frac{\pi}{4}\Bigr)^{\frac{d}{2}}\sum_{n=1}^\infty \frac{1}{n^2}\,{\rm e}^{-\frac{m^2}{eE_0}n\pi} \\
 \times\sum_{k=0}^{\infty}(-1)^k (k+1)\frac{\Gamma(\tfrac{d}{2}+k)}{\Gamma(\tfrac{d}{2})}\Bigl(\frac{eE_0}{m^2}\frac{1}{n\pi}\Bigr)^{\frac{d}{2}+k} \,. \label{eq:ImGammaHE(x)asympt}
\end{multline}
Note that $\lim_{d\to0}\frac{\Gamma(\frac{d}{2}+k)}{\Gamma(\frac{d}{2})}=\delta_{k0}$, where $\delta_{ij}$ is the Kronecker delta, such that \Eqref{eq:ImL1} is recovered for $d=0$.

In analogy to the vacuum decay rate in constant fields, $w(E)=2\,{\rm Im}\{{\cal L}_\text{HE}(E)\}$, we define the vacuum decay rate in the field inhomogeneity~\eqref{eq:E(x)} per space-time volume in the $4-d$ homogeneous directions $V_{(4-d)}$ as
\begin{equation}
 w\bigl[E(x)\bigr]:=\frac{2\,{\rm Im}\bigl\{\Gamma_\text{HE}\bigl[E(x)\bigr]\bigr\}}{V_{(4-d)}}\,. \label{eq:w}
\end{equation}
It is worth emphasizing here that the vacuum decay rate and the pair production rate are not quite the same; cf. \cite{Cohen:2008wz} for a detailed discussion.
Correspondingly, the LCFA one-loop vacuum decay rate in the field inhomogeneity~\eqref{eq:E(x)} with $d\geq0$ can be expressed as
\begin{multline}
 w^{1\text{-loop}}\bigl[E(x)\bigr]=\biggl(\prod_{i=1}^d w_i\biggr)\frac{(eE_0)^2}{4\pi^3}\Bigl(\frac{\pi}{4}\Bigr)^{\frac{d}{2}}\sum_{n=1}^\infty \frac{1}{n^2}\biggl[\Gamma\Bigl(2-\frac{d}{2},\frac{m^2}{eE_0}n\pi\Bigr)-\frac{m^2}{eE_0}n\pi\,\Gamma\Bigl(1-\frac{d}{2},\frac{m^2}{eE_0}n\pi\Bigr)\biggr] \\
 =\biggl(\prod_{i=1}^d w_i\biggr)\frac{(eE_0)^2}{4\pi^3}\Bigl(\frac{\pi}{4}\Bigr)^{\frac{d}{2}}\sum_{n=1}^\infty \frac{1}{n^2}\,{\rm e}^{-\frac{m^2}{eE_0}n\pi}
 \sum_{k=0}^{\infty}(-1)^k (k+1)\frac{\Gamma(\tfrac{d}{2}+k)}{\Gamma(\tfrac{d}{2})}\Bigl(\frac{eE_0}{m^2}\frac{1}{n\pi}\Bigr)^{\frac{d}{2}+k} \,. \label{eq:w(x)}
\end{multline}
With regard to \Eqref{eq:w(x)}, let us in particular highlight the $d$ dependent scaling of the prefactor of the Schwinger exponential ${\rm e}^{-\frac{m^2}{eE_0}n\pi}$ with $(\frac{eE_0}{m^2}\frac{1}{n\pi})^{\frac{d}{2}}$.
Let us emphasize once more that vacuum decay towards a state with real electrons and positrons is only possible if the additional constraint~\eqref{eq:constraint} is also fulfilled.

Especially for weak electric fields $eE_0\ll m^2$, the sums over $n$ and $k$ in the last line of \Eqref{eq:w(x)} can be approximated by their first terms, resulting in
\begin{equation}
 w^{1\text{-loop}}\bigl[E(x)\bigr]\approx\biggl(\prod_{i=1}^d w_i\biggr)\frac{(eE_0)^2}{4\pi^3}\,{\rm e}^{-\frac{m^2}{eE_0}\pi}\Bigl(\frac{eE_0}{m^2}\frac{1}{4}\Bigr)^{\frac{d}{2}} \,. \label{eq:w(x)weak}
\end{equation}
This expression is expected to allow for trustworthy results as long as $\gamma\ll1$ (cf. the discussion above).
Hence, for a one-dimensional ($d=1$) Lorentz profile of the electric field, we find that the imaginary part of the LCFA effective action in the weak field limit and -- by construction -- for $\gamma\ll1$ scales as $\sim(eE_0)^{5/2}\,{\rm e}^{-\frac{m^2}{eE_0}\pi}$. This agrees with the scaling obtained by a semiclassical approximation for the nonperturbative imaginary part of the one-loop effective action for a generic, time-dependent electric field background in this limit \cite{Dunne:2006st}. It had previously been obtained for a time dependent electric field of profile $E(t)=E_0\,{\rm sech}^2(t/\tau)$ by a WKB analysis \cite{Marinov:1977gq}, and also from a Borel resummation of the zeroth-order derivative expansion of the 1-loop effective action in this field  \cite{Dunne:1999uy,Dunne:2000up}. The one-loop Heisenberg-Euler effective action in the latter field profile is known exactly;
cf. \cite{Cangemi:1994by,Cangemi:1995ee,Dunne:1998ni,Dunne:1997kw,Kim:2009pg}, and \cite{Dunne:2004nc,Dunne:2008kc} for reviews.
For a numerical study of pair production in inhomogeneous fields with worldline numerics, cf. \cite{Gies:2005bz}, and for recent pair production studies in inhomogeneous electric fields with worldline instanton methods, cf.~\cite{Dumlu:2011cc,Ilderton:2015qda}.

\section{Conclusions}

In this article we have shown that the LCFA of the Heisenberg-Euler effective action in the specific class of electric fields~\eqref{eq:E(x)} with $0\leq d\leq4$ inhomogeneous directions can be expressed in terms of a single parameter integral over $u\in\mathbb{R}^+$,
with the constant field Lagrangian ${\cal L}^{1\text{-loop}}_{\rm HE}(E)$ with shifted argument, $E\to\frac{E_0}{1+u}$, as integration kernel.
Our results are of special interest as so-far no exact analytical insights into the Heisenberg-Euler action in higher dimensional field inhomogeneities beyond $d=1$ are known.

As an particularly interesting application, we have used these findings to study the imaginary part of the Heisenberg-Euler action in the slowly varying electric field~\eqref{eq:E(x)} as a function of the dimension $d$ of the field inhomogeneity.
The imaginary part of the Heisenberg-Euler action is directly related to the instantaneous vacuum decay rate in the external electric field towards a state with real electrons and positrons and, thus, contains information about the process of electron-positron creation from the quantum vacuum \cite{Schwinger:1951nm}.
By construction, in the limit of $d\to0$, which amounts to a zero-dimensional inhomogeneity, or equivalently, a constant field, we recover the renowned Schwinger formula.
For more dimensional inhomogeneities and peak field strengths $E_0\lesssim E_{\rm cr}$ we generically find an overall parametric suppression of the LCFA instantaneous vacuum decay rate with $\sim(E_0/E_{\rm cr})^{d/2}$.

Finally, we remark that a similar analysis as performed here is, of course, also possible for scalar QED \cite{Weisskopf}.
More specifically, as the one-loop effective Lagrangians of both spinor and scalar QED exhibit the same functional dependence on the electric field amplitude, the identity~\eqref{eq:GammaHE(x)} also holds for scalar QED.
This is obvious, e.g., from a comparison of Eqs.~(1.23)-(1.25) of \cite{Dunne:2004nc} for spinor QED with the analogous expressions for scalar QED, Eqs.~(1.35)-(1.37)  of \cite{Dunne:2004nc}.
Besides, analogous considerations are possible for the photon polarization tensor \cite{Karbstein:2015cpa} in slowly varying electric/magnetic fields of the form \Eqref{eq:E(x)}; cf. \cite{Gies:2013yxa,Nico:2017,IchundNico:2017}.

\acknowledgments

I acknowledge support by the BMBF under grant No. 05P15SJFAA (FAIR-APPA-SPARC). Moreover, I am grateful to H.~Gies for helpful comments.

\appendix

\section{Calculational details}\label{app:id}

Equation~\eqref{eq:Erep} can be straightforwardly derived with the help of the identity
\begin{equation}
 \frac{1}{M^{\frac{n}{2}}}=\frac{1}{\Gamma(\frac{n}{2})}\int_0^\infty{\rm d}\beta\,\beta^{\frac{n}{2}-1}\,{\rm e}^{-\beta M} , \label{eq:M}
\end{equation}
for $\Re\{M\}>0$ and $n\in\mathbb{N}^+$.
First, we use it to obtain
\begin{align}
 \int{\rm d}^4x\,E^2(x)\,{\rm e}^{-{\rm i}\frac{m^2}{eE(x)}s}&=E_0^2\int_0^\infty{\rm d}T\,T\int{\rm d}^4x\,{\rm e}^{-({\rm i}\frac{m^2}{eE_0}s+T)[1+\sum_{i=1}^d(\frac{2x_i}{w_i})^2]} \nonumber\\
 &=V_{(4-d)}\biggl(\prod_{i=1}^d w_i\biggr)\Bigl(\frac{\pi}{4}\Bigr)^{\frac{d}{2}}E_0^2\int_0^\infty{\rm d}T\,\frac{T}{({\rm i}\frac{m^2}{eE_0}s+T)^{\frac{d}{2}}}\,{\rm e}^{-({\rm i}\frac{m^2}{eE_0}s+T)} , \label{eq:s1}
\end{align}
where $V_{(4-d)}:=\int{\rm d}^{4-d}x$, and $0\leq d\leq 4$.
Second, we employ it again to rewrite the integral in the second line of \Eqref{eq:s1} for $d>0$ as
\begin{align}
 \int_0^\infty{\rm d}T\,\frac{T}{({\rm i}\frac{m^2}{eE_0}s+T)^{\frac{d}{2}}}\,{\rm e}^{-({\rm i}\frac{m^2}{eE_0}s+T)}
 &=\frac{1}{\Gamma(\frac{d}{2})}\int_0^\infty\frac{{\rm d}u}{u}\,u^{\frac{d}{2}}\,{\rm e}^{-{\rm i}m^2\frac{1+u}{eE_0}s}\int_0^\infty{\rm d}T\,T\,{\rm e}^{-(1+u)T} \nonumber\\
 &=\frac{1}{\Gamma(\frac{d}{2})}\int_0^\infty\frac{{\rm d}u}{u}\,u^{\frac{d}{2}}\,\frac{1}{(1+u)^2}\,{\rm e}^{-{\rm i}(1+u)\frac{m^2}{eE_0}s} . \label{eq:s2}
\end{align}
Plugging this expression into \Eqref{eq:s1} yields \Eqref{eq:Erep}.

Also note the following identity,
\begin{equation}
 \int{\rm d}^4x\,E^{2n}(x)=V_{(4-d)}\biggl(\prod_{i=1}^d w_i\biggr)\Bigl(\frac{\pi}{4}\Bigr)^{\frac{d}{2}}\,E_0^{2n}\,\frac{\Gamma(2n-\frac{d}{2})}{\Gamma(2n)} \,, \label{eq:IntE2n(x)}
\end{equation}
for $0\leq d<4n$, which can be derived straightforwardly with the help of \Eqref{eq:M}.
The divergence encountered in \Eqref{eq:IntE2n(x)} for $n=1$ and $d\to4$ reflects the fact that the field profile~\eqref{eq:E(x)} is no longer square-integrable in $d=4$ dimensions.

Moreover, for $d>0$ and $\sum_{i=1}^d(\frac{e_E^i}{w_i})^2\neq0$ we have
\begin{equation}
 \int_{-\infty}^\infty{\rm d}\ell\,E(\ell e_E)=\frac{\pi}{2} E_0\frac{1}{\sqrt{\sum_{i=1}^d(\frac{e_E^i}{w_i})^2}}=:E_0 l_E \,, \label{eq:IntE(x)}
\end{equation}
where $l_E$ can be considered as the effective length of the inhomogeneity along the electric field.
On the other hand, for $d=0$ or $\sum_{i=1}^d(\frac{e_E^i}{w_i})^2=0$ we have $l_E\to\infty$.

\end{document}